\documentclass[twocolumn,showpacs,floatfix]{revtex4}

\usepackage{graphicx}

\begin{document}

\newlength{\plotwidth}          
\setlength{\plotwidth}{7.5cm}


\title{\large Spin Blockade in Capacitively Coupled Quantum Dots}

\author{M. C. Rogge}
\email{rogge@nano.uni-hannover.de}
\author{C. F\"uhner}
\author{U.~F. Keyser}
\author{R.~J. Haug}
\affiliation{Institut f\"ur Festk\"orperphysik, Universit\"at
Hannover, Appelstr. 2, 30167 Hannover, Germany}

\date{\today}

\begin{abstract}
We present transport measurements on a lateral double dot produced
by combining local anodic oxidation and electron beam lithography.
We investigate the tunability of our device and demonstrate, that
we can switch between capacitive and tunnel coupling. In the
regime of capacitive coupling we observe the phenomenon of spin
blockade in a magnetic field and analyze the influence of
capacitive interdot coupling on this effect.
\end{abstract}

\maketitle

In recent decades great progress has been made in the development
and measurement of quantum dot devices \cite{Kouwenhoven-97}. Many
single electron transistors have been fabricated and during the
last years coupled quantum dots (e.g.
\cite{Wiel-03,Holleitner-02,Livermore-96,Hofmann-95,Molenkamp-95})
came more and more into the focus of research, as they are
proposed as crucial parts for quantum computers \cite{Loss-98}.
Since preparation and detection of electron spin is another
essential element in future quantum information applications, the
discovery of spin blockade in quantum dot devices was a great step
forward \cite{Ciorga-00,Ciorga-02b}. It has been observed in
single quantum dots defined by metallic gates. Coupled quantum
dots are also of great interest in the regime of spin blockade.
Only recently spin resolved measurements on molecular states of
two dots tunnel coupled in series have been made using spin
blockade \cite{Pioro-Ladriere-03}.


In this letter we present our results for spin blockade in a
parallel double quantum dot based on local anodic oxidation (LAO).
Due to the complexity of the device we first have to characterize
our system in terms of the interdot coupling, which depends on top
gate voltage and magnetic field. We find, that the relevant
regime, where we observe spin blockade, features capacitive
coupling only. We show our measurements for spin blockade and
analyze the influence of capacitive interdot coupling on this
effect.

Our quantum dot device is based on a GaAs/AlGaAs heterostructure
with a two-dimensional electron system (2DES) 34~nm below the
surface. The sheet density is $n=4.3\cdot 10^{15}$~m$^{-2}$. We
apply two different nanolithographic techniques to the sample
surface to define two adjacent quantum dots connected in parallel
to common source and drain contacts. We use an atomic force
microscope (AFM) to write the basic double dot structure by local
anodic oxidation. We complete the structure with a metallic gate
patterned with electron beam lithography (e-beam) to add the
function of controlled tunability of the interdot coupling.

\begin{figure}
 \includegraphics{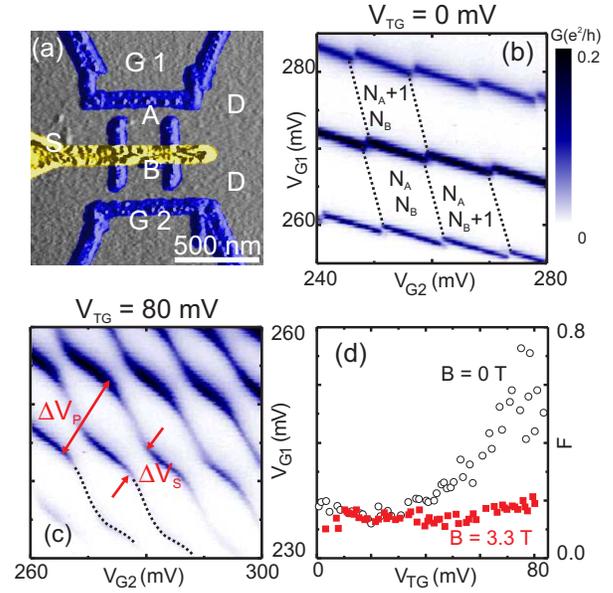}
 \caption{(a) Colorized AFM image of our double dot device.
Oxide lines (blue) define source (S) and drain (D) leads, side
gates G1 and G2 and two quantum dots A and B. A metallic top gate
(golden) is added by e-beam lithography. (b) $G$ as a function of
both side gate voltages $V_{G1}$ and $V_{G2}$ for $V_{TG}$=0~mV.
The Coulomb Blockade of dot A is interrupted by regular splittings
referring to transitions on dot B. (c) Similar plot for
$V_{TG}$=80~mV. The splitting has increased, round edges and
additional features completing the hexagonal pattern represent
molecular binding states.(d) Fractional peak splitting $F$ as a
function of top gate voltage $V_{TG}$ for $B$=0~T and $B$=3.3~T.}
 \label{fig1}
\end{figure}
An AFM image of our device is given in Fig. \ref{fig1}(a). Those
parts, that have been fabricated by LAO, are colored blue. They
represent insulating lines, that form a complete double dot device
including the two dots A and B, side gates G1 and G2 and source
(S) and drain (D) leads. The two dots are connected via 80 to
100~nm wide tunneling barriers to the leads and via a central
barrier to each other. An additional metallic top gate (golden)
can be biased to tune the height of the central barrier and thus
to tune the interdot coupling. A detailed description of a similar
device and the fabrication technique can be found in
\cite{Rogge-03}.

Our double dot device is characterized in transport measurements
in a $^3$He/$^4$He dilution refrigerator at a base temperature of
70~mK. We apply standard Lock In technique to measure the
differential conductance $G$ through the double dot system. To
analyze the features for spin blockade, which will be shown later,
it is first necessary to understand the interaction of both dots
in the relevant parameter space. Therefore we start investigating
the interdot coupling with the top gate grounded at $B$=0~T.
Figure \ref{fig1}(c) shows $G$ under these conditions as a
function of both side gate voltages $V_{G1}$ and $V_{G2}$. The
dark regions denote high values for $G$ and represent transport
over dot A. No lines corresponding to transport over dot B are
visible meaning that B is not connected to both leads.
Nevertheless the equidistant splittings interrupting the features
of dot A indicate the presence of dot B, which is connected to the
source lead while the drain lead remains closed for all the
parameters used here, as we know from nonlinear measurements (not
shown here). This is due to the lithographic gap width of only
70~nm for the barrier between dot B and drain compared to 80 to
100~nm for the other barriers. Thus we can complete the observed
features by dotted lines to a hexagonal pattern typical for two
coupled quantum dots. Crossing a dark line changes the electron
number on dot A ($N_A$), crossing a dotted line changes the charge
on dot B ($N_B$). Due to the sharp corners of the hexagons and the
fact that the only features visible are those of dot A, the
interdot coupling in this regime is purely capacitive.

Figure \ref{fig1}(d) shows a similar plot but with the top gate
voltage set to 80~mV. Due to the reduced height of the central
barrier at $V_{TG}$=80~mV the interdot coupling has increased
leading to a very wide splitting. In contrast to Fig.
\ref{fig1}(c) the edges of the hexagonal pattern are clearly
rounded and there are finite conductance values visible for
transitions on dot B. Both effects demonstrate the molecular like
character of the system. Instead of atomic transitions the visible
features refer to molecular binding states indicating tunnel
coupling \cite{Blick-96,Blick-98}.

A more detailed view can be gained by calculating the fractional
peak splitting $F=2\Delta V_S/\Delta V_P$ (see Fig.
\ref{fig1}(c)), which is 0 for two separate dots and 1 for two
totally merged dots \cite{Golden-96}. Figure \ref{fig1}(d) shows
$F$ as a function of top gate voltage for $B$=0~T (open circles).
For voltages below 30~mV the fractional peak splitting is
unaffected by $V_{TG}$ and remains at values below 0.2. This is
due to capacitive coupling, that depends on the LAO geometry of
both dots, which is almost not affected by top gate voltage. At 30
to 40~mV tunnel coupling sets in and molecular features appear.
With increasing top gate voltage the fractional peak splitting can
be tuned to almost 0.8 corresponding to very strong tunnel
coupling.

The interdot coupling is not only changed by top gate voltage but
also by a magnetic field applied perpendicular to the 2DES. At
$B$=3.3~T the fractional peak splitting is again decreased to a
value below 0.2 and remains unaffected by top gate voltage
(squares in Fig. \ref{fig1}(d)). The magnetic field destroys the
tunnel coupling and moves the system back to the regime of
capacitive coupling even with $V_{TG}$=80~mV.

\begin{figure}
 \includegraphics{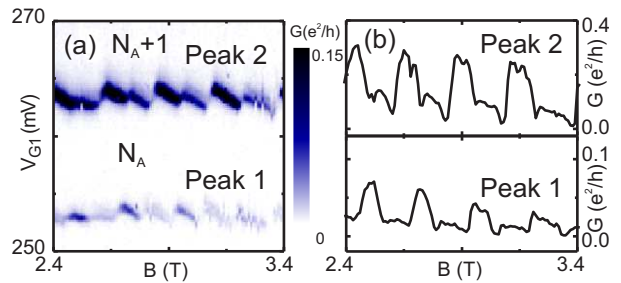}
 \caption{(a) $G$ as a function of $V_{G1}$ and magnetic field. The two
Coulomb peaks of dot A show oscillations in peak position in the
regime of spin blockade. (b) The maxima for both peaks as a
function of $B$ show a spin dependent height, high values for spin
down transport, low values for spin up.}
 \label{fig2}
\end{figure}

After having characterized our double dot we can now focus on the
features of spin blockade, that appear around 3.3~T in the regime
of capacitive coupling.

Figure \ref{fig2}(a) shows $G$ as a function of $V_{G1}$ and $B$.
Two Coulomb peaks are visible denoting transitions on dot A. For
both peaks oscillations in peak position are obvious as well as
alternating amplitudes for the peak maxima (see Fig.
\ref{fig2}(b)). The amplitude is changed by magnetic field and by
electron number $N_{A}$. Therefore this effect looks similar to
the effect of spin blockade that has been observed in several
single dot devices \cite{Ciorga-00,Ciorga-02b}.

Spin blockade is explained as follows. As source and drain consist
of a 2DES, they develop edge channels in a magnetic field. If the
potential is flat enough these channels are separated in space
leaving a spin down channel of the lowest Landau level near to the
dot and a spin up channel further away. Thus the overlap of
electronic wave functions of leads and dot is spin dependent. Spin
up electrons couple weaker to the dot than spin down electrons.
Therefore transport with spin up electrons is suppressed. A
magnetic field changes the available spin state for electrons
tunneling into the dot \cite{McEuen-92}, which becomes obvious in
alternating peak positions and peak amplitudes of Coulomb blockade
peaks. LAO devices are known to have steeper potentials than split
gate devices \cite{Fuhrer-01,Held-99}. Thus the spatial separation
of edge channels in the leads is believed to be much smaller in
LAO devices. Spin blockade should be very improbable or at least
suppressed in comparison to split-gate devices and has actually
not been measured in LAO devices so far.

Nevertheless, despite the LAO nature we observe these features
here very clearly, and although this is a coupled dot system, we
assume, that this observation is due to spin blockade in dot A
only, since we are in the regime of capacitive coupling.
Nevertheless the influence of dot B, which is still present, must
be investigated. Therefore we measured $G$ as a function of
magnetic field and top gate voltage. In contrast to side gate 1,
that couples mainly to dot A, the top gate has a stronger
influence on dot B. Thus we can expect to see features from both
dots at once.
\begin{figure}
 \includegraphics[scale=1.5]{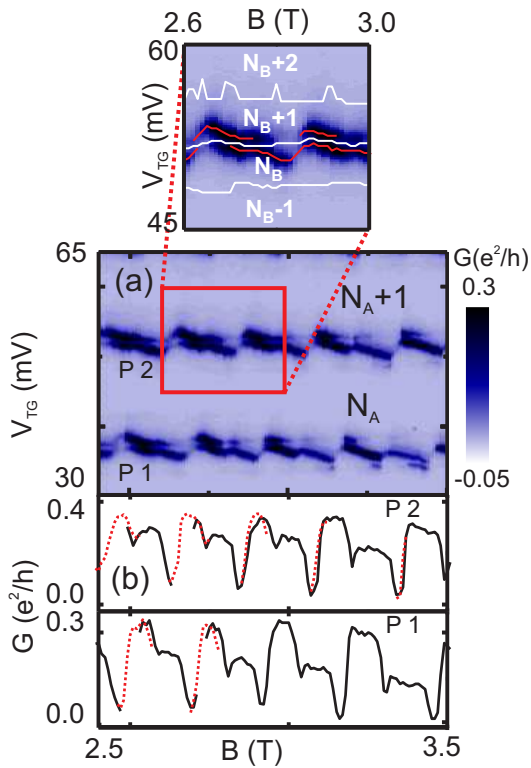}
 \caption{(a) $G$ as a function of $V_{TG}$ and magnetic field. The peaks of
dot A (red lines in the section) show oscillating positions. Due
to capacitive coupling these peaks are regularly split by
transitions on dot B (white lines). Thus each peak occurs below
and above such a splitting. (b) The maxima for both peaks show
again an alternating height caused by single dot spin blockade.
The amplitudes for parts of a peak below a splitting (solid) and
for parts above a splitting (dotted) fit almost perfectly
together. Thus capacitive dot-dot coupling has no effect on spin
blockade.}
 \label{fig3}
\end{figure}

The measurement is shown in Fig. \ref{fig3}(a). Again we observe
oscillating peak positions in the Coulomb peaks of dot A. But a
detailed look reveals, that these peaks are split by local minima
in differential conductance. These minima are colored white in the
highlighted section and show a pattern of more or less horizontal
lines with regular distance in $V_{TG}$. These lines refer to
transitions on dot B and correspond to the splitting of triple
points in Fig. \ref{fig1}(b). As the Coulomb peaks of dot A are
split, parts of each peak are above a splitting and others are
below (red lines in the section). Therefore we can compare the
peak amplitudes of each Coulomb peak below and above a splitting
and can directly investigate the effect of capacitive dot-dot
coupling on the peak amplitude. This is done in Fig.
\ref{fig3}(b). For each peak visible in Fig. \ref{fig3}(a) the
peak amplitude is plotted as a function of magnetic field. For all
peaks both amplitudes below a splitting (solid line) and above a
splitting (dotted line) fit almost perfectly together and the
typical oscillations of high and low amplitude become visible.
Thus the influence of capacitive interdot coupling is identified
as a pure peak splitting effect without any impact on the peak
amplitude. In consequence we find our assumption verified, that
the oscillations in peak position and amplitude ascribe to a
single dot spin blockade of dot A. As a result this spin blockade
is not destroyed by capacitive interdot coupling, it is not even
affected at all.

In summary we have investigated spin blockade in a tunable double
dot device fabricated by LAO. We have characterized the system
regarding the tunability of the interdot coupling as a function of
top gate voltage and magnetic field. In the regime of capacitive
coupling we observed single dot spin blockade. As far as we know
this is the first time that this has been measured in devices
fabricated by LAO. We have shown that this spin blockade is not
destroyed or at least affected by transitions on the capacitively
coupled dot. Therefore it is a very useful tool for spin detection
even in LAO based quantum dot devices.

We thank M. Bichler, G. Abstreiter and W. Wegscheider for the
heterostructure and D. Pfannkuche for helpful discussions. We
thank F. Hohls for careful reading of the manuscript. This work
has been supported by BMBF.


\newpage




\end{document}